\documentclass{llncs}
\usepackage[utf8]{inputenc}
\usepackage[T1]{fontenc}
\usepackage{graphicx}
\usepackage{amsmath}
\usepackage{cite}
\usepackage{multirow}
\usepackage{xcolor}
\usepackage{hyperref}
\usepackage{tabularx}

\newcommand{\ie}{\emph{i.e.},~}
\newcommand{\eg}{\emph{e.g.},~}

\title{Deep Grading based on Collective Artificial Intelligence for AD Diagnosis and Prognosis}
\author{Huy-Dung Nguyen \inst{1} \and Michaël Clément \inst{1} \and Boris Mansencal \inst{1} \and Pierrick Coupé \inst{1}}
\institute{~\inst{1}Univ. Bordeaux, CNRS, Bordeaux INP, LaBRI, UMR 5800, 33400 Talence, France}
\date{June 2021}

\begin{document}

\maketitle

\begin{abstract}

Accurate diagnosis and prognosis of Alzheimer's disease are crucial to develop new therapies and reduce the associated costs. Recently, with the advances of convolutional neural networks, methods have been proposed to automate these two tasks using structural MRI.
However, these methods often suffer from lack of interpretability, generalization, and can be limited in terms of performance. In this paper, we propose a novel deep framework designed to overcome these limitations. Our framework consists of two stages. In the first stage, we propose a deep grading model to extract meaningful features. To enhance the robustness of these features against domain shift, we introduce an innovative collective artificial intelligence strategy for training and evaluating steps. In the second stage, we use a graph convolutional neural network to better capture AD signatures. Our experiments based on 2074 subjects show the competitive performance of our deep framework compared to state-of-the-art methods on different datasets for both AD diagnosis and prognosis.

\end{abstract}

\section{Introduction}

%Alzheimer's Disease (AD) is a common neurodegenerative disease characterized by the progressive impairment of cognitive functions and the most common cause of dementia. In 2006, there were 26.6 million AD patients worldwide~\cite{brookmeyer_adinfo}, increasing to 46.8 million in 2015 and expected to reach 131.5 million in 2050~\cite{worldalzheimerreport2016}. The cost of carrying for AD patients is also expected to rise dramatically depending on the disease severity. Thus, early detection of Alzheimer's Disease is highly desirable for developing new therapies, slowing down disease progression and consequently reducing the associated costs.

The first cognitive symptoms of Alzheimer's disease (AD) appear right after the morphological changes caused by brain atrophy~\cite{adsign}. Those changes can be identified with the help of structural magnetic resonance imaging (sMRI)~\cite{BRON2015562}. Recently, with the advances of convolutional neural networks (CNN), methods have been proposed for automatic AD diagnosis using sMRI. Despite encouraging results, current deep learning methods suffer from several limitations. First, deep models lack transparency in their decision-making process~\cite{Eduardo_explainable_deep_cnn, Zhang_explainable_res_self}. Therefore, this limits their use for computer-aided diagnosis tools in clinical practice. Second, for medical applications, the generalization capacity of classification models is essential. However, only a few works have proposed methods robust to domain shift~\cite{wachinger_DomainAdaptation, ehsan_3D_CNN_Adaptation}. Third, current CNN models proposed for AD diagnosis and prognosis still perform poorly~\cite{WEN2020101694}. Indeed, when properly validated on external datasets, current CNN-based methods perform worse than traditional approaches (\ie standard linear SVM).

\begin{figure}[t]
\centering
\includegraphics[width=\textwidth]{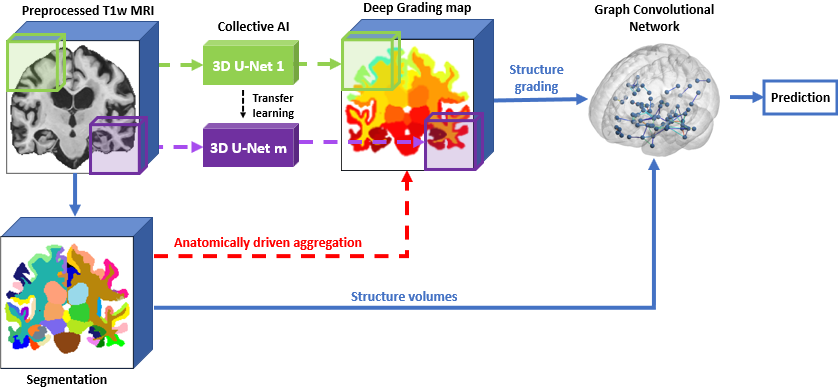}
\caption{Overview of our processing pipeline. The MRI image, its segmentation and the deep grading map illustrated are from an AD subject.}
\label{fig:pipeline}
\end{figure}

In this paper, to address these three major limitations, we propose a novel interpretable, generalizable and accurate deep framework.
An overview of our proposed pipeline is shown in Figure~\ref{fig:pipeline}.
First, we propose a novel Deep Grading (DG) biomarker to improve the interpretability of deep model outputs. Inspired by the patch-based grading frameworks~\cite{tontong_grading, hett_grading, COUPE2012141}, this new biomarker provides a grading map with a score between $-1$ and $1$ at each voxel related to the alteration severity. This interpretable biomarker may help clinicians in their decision and to improve our knowledge on AD progression over the brain.
Second, we propose an innovative collective artificial intelligence strategy to improve the generalization across domains and to unseen tasks. As recently shown for segmentation~\cite{coupe2020assembly, kamraoui2020broader}, the use of a large number of networks capable of communicating offers a better capacity for generalization. Based on a large number of CNNs (\ie 125 U-Nets), we propose a framework using collective artificial intelligence efficient on different datasets and able to provide accurate prognosis while trained for diagnosis task.
Finally, we propose to use a graph-based modeling to better capture AD signature using both inter-subject similarity and intrasubject variability. As shown in~\cite{hett_grading}, such strategy improves performance in AD diagnosis and prognosis.

In this paper, our main contributions are threefold:
\begin{itemize}
    \item A novel deep grading biomarker providing interpretable grading maps.
    \item An innovative collective artificial intelligence strategy robust to unseen datasets and unknown tasks.
    \item A new graph convolutional network (GCN) model for classification offering state-of-the-art performance for both AD diagnosis and prognosis.
\end{itemize}

\section{Materials and method}

\subsection{Datasets}

The data used in this study, consisting of 2074 subjects, were obtained from multiple cohorts: the Alzheimer's Disease Neuroimaging Initiative (ADNI)~\cite{adni_dataset}, the Open Access Series of Imaging Studies (OASIS)~\cite{oasis3_dataset}, the Australian Imaging, Biomarkers and Lifestyle (AIBL)~\cite{aibl_dataset}, the Minimal Interval Resonance Imaging in Alzheimer's Disease (MIRIAD)~\cite{miriad_dataset}. We used the baseline T1-weighted MRI available in each of these studies. Each dataset contains AD patients and cognitively normal (CN) subjects. For ADNI1 and AIBL, it also includes mild cognitive impairment (MCI), the early stage of AD composed of abnormal memory dysfunctions. Two groups of MCI are considered: progressive MCI (pMCI) and stable MCI (sMCI). The definition of these two groups is the same as in~\cite{WEN2020101694}. Table~\ref{tab:datasets} summarizes the number of participants for each dataset used in this study. During experiments, AD and CN from ADNI1 are used as training set and the other subjects as testing set.

\begin{table}[t]
    \centering
    \fontsize{8}{10}\selectfont
    \caption{Number of participants used in our study. Data used for training is in bold.}
    \label{tab:datasets}
    \begin{tabular}{l@{\hskip 3em}c@{\hskip 2em}c@{\hskip 2em}c@{\hskip 2em}c}
        \hline
        Dataset      & CN         & AD        & sMCI  & pMCI \\
        \hline
        \textbf{ADNI1}        &  \textbf{170}       & \textbf{170}       & 129   & 171 \\
        ADNI2        &  149       & 149       & -     & - \\
        AIBL         &  233       & 47        & 12    & 20 \\
        OASIS3       &  658       & 97        & -     & - \\
        MIRIAD       &  23        & 46        & -     & - \\
        \hline
    \end{tabular}
\end{table}
 
\subsection{Preprocessing}

All the T1w MRI are preprocessed using the following steps: (1) denoising~\cite{denoising}, (2) inhomogeneity correction~\cite{inhomogeneity_correction}, (3) affine registration into MNI space ($181\times217\times181$ voxels at $1\text{mm}\times1\text{mm}\times1\text{mm}$)~\cite{affine_registration}, (4) intensity standardization~\cite{intensity_normalisation} and (5) intracranial cavity (ICC) extraction~\cite{brain_extraction}. After that preprocessing, we use AssemblyNet~\cite{coupe2020assembly} to segment 133 brain structures (see Figure~\ref{fig:pipeline}). The list of structures is the same as in~\cite{Yuankai_slant}. In this study, brain structure segmentation is used to determine the structure volume (\ie normalized volume in \% of ICC) and aggregate information in the grading map (see Section~\ref{section:deep_grading} and Figure~\ref{fig:pipeline}). 

\subsection{Deep Grading for disease visualization}
\label{section:deep_grading}

In AD classification, most of deep learning models only use CNN as binary classification tool. In this study, we propose to use CNN to produce 3D maps indicating where specific anatomical patterns are present and the importance of structural changes caused by AD.

To capture these anatomical alterations, we extend the idea of the patch-based grading (PBG) framework~\cite{COUPE2012141, hett_grading, tontong_grading}. The PBG framework provides a 3D grading map with a score between $-1$ and $1$ at each voxel related to the alteration severity. Contrary to previous PBG methods based on non-local mean strategy, here we propose a novel DG framework based on 3D U-Nets.

Concretely, each U-Net (similar to~\cite{coupe2020assembly}) takes a 3D sMRI patch (\eg $32 \times 48 \times 32$) and outputs a grading map with values in range $[-1, 1]$ for each voxel. Voxels with a higher value are considered closer to AD, while voxels with a lower value are considered closer to CN. For the ground-truth used during training, we assign the value $1$ (resp. $-1$) to all voxels inside a patch extracted from an AD patient (resp. CN subject). All voxels outside of ICC are set to~$0$.

Once trained, the deep models are used to grade patches. These local outputs are gathered to reconstruct the final grading map (see Section~\ref{section:collective_AI}). Using the structure segmentation, we represent each brain structure grading by its average grading score (see Figure~\ref{fig:pipeline}). This anatomically driven aggregation allows better and meaningful visualization of the disease progression. In this way, during the classification step (see Section~\ref{section:GCN}), each subject is encoded by an n-dimensional vector where n is the number of brain structures.

\subsection{Collective AI for grading}
\label{section:collective_AI}

As recently shown in~\cite{cross_cohort_bron, WEN2020101694}, current AD classification techniques suffer from a lack of generalization. In this work, we propose an innovative collective artificial intelligence strategy to improve the generalization across domains and to unseen tasks. As recently shown for segmentation~\cite{coupe2020assembly, kamraoui2020broader}, the use of a large number of compact networks capable of communicating offers a better capacity for generalization. There are many advantages to using the collective AI strategy. First, it addresses the problem of GPU memory in 3D since each model processes only a sub-volume of the image. The use of a large number of compact networks is equivalent to a big neural network with more filters. Second, the voting system based on a large number of specialized and diversified models helps the final grading decision to be more robust against domain shift and different tasks.

Concretely, a preprocessed sMRI is decomposed into $k\times k \times k$ overlapping patches of the same size (\eg $32 \times 48 \times 32$). During training, for each patch localization in the MNI space, a specialized model is trained. Therefore, in our case ($k=5$), we trained $m=k\times k\times k = 125$ U-Nets to cover the whole image (see Fig. 1). Moreover, each U-Net is initialized using transfer learning from its nearest neighbor U-Nets in the MNI space, except the first one trained from scratch as proposed in~\cite{coupe2020assembly}. As adjacent patches have some common patterns, this communication allows grading models to share useful knowledge between them. For each patch, 80\% of the training dataset (\ie ADNI1) is used for training and the remaining 20\% for validation. The accuracy obtained on validation set is used to reconstruct the final grading map using a weighted average as follows:

\begin{equation}
    \label{eq:weighted_patch}
    \begin{aligned}
        G_i = \frac{\sum_{x_i \in P_j}{\alpha_j} * g_{ij}}{\sum_{x_i \in P_j}{\alpha_j}}
    \end{aligned}
\end{equation}
where $G_i$ is the grading score of the voxel $x_i$ in the final grading map, $g_{ij}$ is the grading score of the voxel $x_i$ in the local grading patch $P_j$, and $\alpha_j$ is the validation accuracy of the patch $j$. This weighted vote enables to give more weight to the decision of accurate models during the reconstruction.

\subsection{Graph convolutional neural network for classification}
\label{section:GCN}

The DG feature provides an inter-subject similarity biomarker which is helpful to detect AD signature. However, the structural alterations leading to cognitive decline could be different between subjects. Indeed, following the idea of~\cite{hett_grading}, we model the intra-subject variabilities by a graph representation to capture the relationships between several regions related to the disease. We define an undirected graph $G = (N, E)$, where $N = \{n_1, \ldots, n_s\}$ is the set of nodes for the $s$ brain structures and $E = s\times s$ is the matrix of edge connections. In our approach, all nodes are connected with each other in a complete graph, where nodes embed brain features (\eg our proposed DG feature) and potentially other types of external features.

Indeed, besides the grading map, the volume of structures obtained from the segmentation could be helpful to distinguish AD patients from CN~\cite{hett_grading, tontong_grading}. It is due to the evidence that AD leads to structure atrophy. Age is also an important factor as, within sMRI, patterns in the brain of young AD patients could be similar to elder CN. Indeed, the combination of those features is expected to improve our classification performance. In our method, each node represents a brain structure and embeds a feature vector (DG, V, A) where V and A are respectively the volume of structures and subject's age. Finally, we use the graph convolutional neural network (GCN)~\cite{kipf2017semi} as the way to pass messages between nodes and to perform final classification.

\subsection{Implementation details}

First, we downsample the sMRI from $181 \times 217 \times 181$ voxels (at 1mm) to $91 \times 109 \times 91$ voxels to reduce the computational cost, then decompose them into $5\times5\times5$ overlapping patches of size $32 \times 48 \times 32$ voxels equally spaced along the three axis. For each patch, an U-Net is trained using mean absolute error loss, Adam optimizer with a learning rate of $0.001$. The training process is stopped after 20 epochs without improvement in validation loss. We employed several data augmentation and sampling strategies to alleviate the overfitting issue during training. A small perturbation is first created in training samples by randomly translating by $t \in \{-1, 0, 1\}$ voxel in 3 dimensions of the image. We then apply the mixup~\cite{mixup} data augmentation scheme that was shown to improve the generalization capacity of CNN in image classification.

Once the DG feature is obtained, we represent each subject by a graph of 133 nodes. Each node represents a brain structure and embeds DG, volume and age features. Our classifier is composed of 3 layers of GCN with 32 channels, followed by a global mean average pooling layer and a fully connected layer with an output size of 1. The model is trained using the binary cross-entropy loss, Adam optimizer with a learning rate of $0.0003$. The training process is stopped after 20 epochs without improvement in validation loss. At inference time, we randomly add noise $X \sim \mathcal{N}(0, 0.01)$ to the node features and compute the average of 3 predictions to get the global decision. Experiments have shown that it helps our GCN to be more stable.

For training and evaluating steps, we use a standard GPU (\ie NVIDIA TITAN~X) with 12Gb of memory.

\section{Experimental results}

In this study, the grading models and classifiers are trained using ADNI1 dataset within AD and CN subjects. Then, we assess their generalization capacity in domain shift using AD, CN subjects from ADNI2, AIBL, OASIS, MIRIAD. The generalization capacity in derived tasks is performed using pMCI, sMCI subjects from ADNI1 (same domain) and AIBL (out of domain).

\textbf{Influence of collective AI strategy.}
In this part, the DG feature is denoted as $DG_C$(resp. $DG_I$) when obtained with the collective (resp. individual) AI strategy. The individual AI strategy refers to the use of a single U-Net to learn patterns from all patches of sMRI. We compare the efficiency of $DG_C$ and $DG_I$ feature when using the same classifier (\ie SVM or GCN) (see Table~\ref{table:method_assessment}). These experiments show that using $DG_C$ achieves better results in most configurations. When using SVM classifier, we observe a gain of 3.6\% (resp. 0.8\%) on average in AD/CN (resp. pMCI/sMCI) classification. The efficiency of $G_C$ feature is even better with GCN classifier, where a gain of 4.0\% (resp. 3.5\%) is observed.

\textbf{Influence of GCN classifier.}
Besides the DG feature, the intra-subject variabilities are also integrated into our graph representation. Hence, it should be beneficial to use GCN to exploit all this information. In our experiments, GCN outperforms SVM in all the tests using either $DG_I$ or $DG_C$ feature (see Table~\ref{table:method_assessment}). Concretely, using $DG_I$ feature, we observe a gain of 5.0\% (resp. 7.6\%) on average for AD/CN (resp. pMCI/sMCI) classification. These improvements are 5.4\% and 10.6\% when using $DG_C$ feature.

\textbf{Influence of using additional non-image features.}
Moreover, we analyze the model performance using $DG_C$ with the structural volume $V$ and age $A$ as additional node features in our graph representation. By using the combined features, the performance on average in AD/CN and pMCI/sMCI is both improved by 0.3\% and 1.4\% compared to $DG_C$ feature (see Table~\ref{table:method_assessment}). In the rest of this paper, these results are used to compare with current methods.

\newcolumntype{Y}{>{\centering\arraybackslash}X}
\begin{table}[t]
    \centering
    \fontsize{8}{10}\selectfont
    \caption{Validation of the collective AI strategy, GCN classifier, the combination of DG feature with other image and non-image features using GCN classifier. {\color{red}Red}: best result, {\color{blue}Blue}: second best result. The balanced accuracy (BACC) is used to assess the model performance. The results are the average accuracy of 10 repetitions and presented in percentage. All the methods are trained on the AD/CN subjects of the ADNI1 dataset.}
    \label{table:method_assessment}
    
    \begin{tabularx}{\textwidth}{c|c|@{}YYYY|YY|Y|Y@{}}

    \hline
        \multirow{2}{*}{Classifier} & \multirow{2}{*}{Features} & \multicolumn{4}{c|}{AD/CN} & \multicolumn{2}{c|}{pMCI/sMCI} & \multicolumn{2}{c}{Average}\\  \cline{3-10}
        & & ADNI2 & OASIS & MIRIAD & AIBL & ADNI1 & AIBL & AD/CN & p/sMCI\\ 
        
    \hline
    SVM & $DG_I$ &
        83 &
        83 &
        88 &
        79 &
        65 &
        66 &
        83.3 &
        65.5\\
        
    SVM & $DG_C$ &
        83 &
        84 &
        91 &
        87 &
        68 &
        64 &
        86.3 &
        66.0 \\
    
    GCN & $DG_I$ &
        \text{\color{blue}84} &
        \text{\color{blue}88} &
        96 &
        82 &
        68 &
        73 &
        87.5 &
        70.5\\
    
    GCN & $DG_C$ &
        \text{\color{red}87} &
        \text{\color{red}89} &
        \text{\color{red}100} &
        \text{\color{blue}88} &
        \text{\color{blue}70} &
        \text{\color{red}76} &
        \text{\color{blue}91.0} &
        \text{\color{blue}73.0}\\
        
    GCN & $DG_C, V, A$ &
        \text{\color{red}87} &
        \text{\color{blue}88} &
        \text{\color{blue}98} &
        \text{\color{red}92} &
        \text{\color{red}74} &
        \text{\color{blue}74} &
        \text{\color{red}91.3} &
        \text{\color{red}74.0}\\
        
    \hline
    \end{tabularx}
\end{table}

\begin{table}[ht!]
    \centering
    \fontsize{8}{10}\selectfont
    \caption{Comparison of our method with current methods in AD diagnosis and prognosis. {\color{red}Red}: best result, {\color{blue}Blue}: second best result. The balanced accuracy (BACC) is used to assess the model performance. All the methods are trained on the AD/CN subject of the ADNI1 dataset (except~\cite{hierachical_fcn} that is fined-tuned on MCI subjects for sMCI/pMCI task).}
    \label{table:Comparison_with_sota}
    
    \begin{tabularx}{\textwidth}{l@{}|YYYY|YY@{}}

    \hline
    \multirow{2}{*}{Methods} &\multicolumn{4}{c|}{AD/CN} & \multicolumn{2}{c}{pMCI/sMCI}\\ \cline{2-7}
    & ADNI2 & OASIS & MIRIAD & AIBL & ADNI1 & AIBL\\
    \cline{1-7}
    %\cline{1-1} \cline{2-7}
    Landmark-based CNN ~\cite{landmark_based} &
        \text{\color{red} 91} &
        - &
        92 &
        - &
        - &
        -\\
        
    %\cline{1-1} \cline{2-7}
    Hierachical FCN ~\cite{hierachical_fcn} &
        \text{\color{blue} 89} &
        - &
        - &
        - &
        69 &
        - \\
    
    Patch-based CNN ~\cite{WEN2020101694} &
        - &
        64 &
        - &
        81 &
        70 &
        64\\

    ROI-based CNN ~\cite{WEN2020101694} &
        - &
        69 &
        - &
        84 &
        70 &
        60 \\

    Subject-based CNN ~\cite{WEN2020101694} &
        - &
        67 &
        - &
        83 &
        69 &
        52 \\

    Voxel-based SVM ~\cite{WEN2020101694} &
        - &
        70 &
        - &
        88 &
        \text{\color{red}75} &
        62\\
    
    AD$^2$A ~\cite{Guan2020AttentionGuidedDD} &
        88 &
        - &
        - &
        88 &
        - &
        - \\

    Efficient 3D ~\cite{efficient_3D_cnn} &
       - &
        \text{\color{red} 92} &
        \text{\color{blue} 96} &
        \text{\color{blue}91} &
        70 &
        \text{\color{blue}65} \\

    3D Inception-ResNet-v2 ~\cite{Lu2020.08.18.256594} &
        - &
        85 &
        - &
        \text{\color{blue}91} &
        42 &
        - \\
    
    \cline{1-1} \cline{2-7}
    
    Our method &
        87 &
        \text{\color{blue}88} &
        \text{\color{red}98} &
        \text{\color{red}92} &
        \text{\color{blue}74} &
        \text{\color{red}74} \\
    \hline
    
    \end{tabularx}
\end{table}

\textbf{Comparison with state-of-the-art methods.}
Table~\ref{table:Comparison_with_sota} summarizes the current performance of state-of-the-art methods proposed for AD diagnosis and prognosis classification that have been validated on external datasets.
In this comparison we considered five categories of deep methods: patch-based strategy based on a single model (Patch-based CNN~\cite{WEN2020101694}), patch-based strategy based on multiple models (Landmark-based CNN~\cite{landmark_based}, Hierarchical FCN~\cite{hierachical_fcn}), ROI-based strategy based on a single model focused on hippocampus (ROI-based CNN~\cite{WEN2020101694}), subject-based considering the whole image based on a single model (subject-based CNN~\cite{WEN2020101694}, 3D Inception-ResNet-v2~\cite{Lu2020.08.18.256594}, Efficient 3D~\cite{efficient_3D_cnn} and  AD$^2$A~\cite{Guan2020AttentionGuidedDD}) and a classical voxel-based model using a SVM (Voxel-based SVM~\cite{WEN2020101694}).

For AD diagnosis (\ie AD/CN), all the methods show good balanced accuracy, although some of them failed to generalize on OASIS. In this scenario (unseen datasets), our method obtained high accuracy for all the datasets.
This confirms the generalization capacity of our approach against domain shift.

For AD prognosis (\ie pMCI/sMCI), we observe a significant drop for all the methods. This drop is expected since pMCI/sMCI classification is more challenging and since models are trained on a different task (\ie AD/CN). For this task, our method is generally robust, especially on AIBL. Moreover, our approach is the only deep learning method that performs competitively with the SVM model~\cite{WEN2020101694} on ADNI1, while significantly better on AIBL. In this scenario (unknown task), our method obtains the highest accuracy on average.
These results highlight the potential performance of our method on unseen tasks.

\begin{figure}[t] % [ht]
\centering
\includegraphics[width=0.8\textwidth]{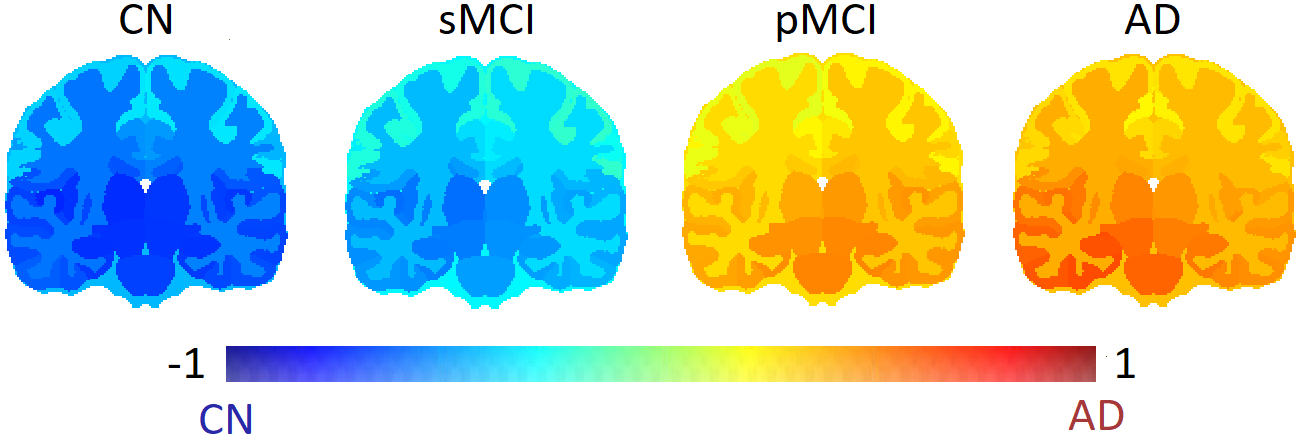}
\caption{Average grading map per group of subjects.}
\label{fig:visualization}
\end{figure}

\textbf{Interpretation of collective deep grading.} % feature
To highlight the interpretability capabilities offered by our DG feature, we first compute the average DG map for each group: AD, pMCI, sMCI and CN (see Figure~\ref{fig:visualization}). First, we can note that the average grading maps increase between each stage of the disease. Second, we estimated the top 10 structures with highest absolute value of grading score over all the testing subjects. The found structures are known to be specifically and early impacted by AD. These structures are:
\textit{bilateral hippocampus}~\cite{Frisoni_clinical_use_sMRI},
\textit{left amygdala} and \textit{left inferior lateral ventricle}~\cite{coupe2019lifespan},
\textit{left parahippocampal gyrus}~\cite{Kesslak51},
\textit{left posterior insula}~\cite{Foundas_insula},
\textit{left thalamus proper}~\cite{jong_thalamus},
\textit{left transverse temporal gyrus}~\cite{Liu2012edu},
\textit{left ventral diencephalon}~\cite{Lebedeva_ventral_dc}.
While other attention-based deep methods failed to find structures related to AD~\cite{cross_cohort_bron}, our DG framework shows high correlation with current physiopathological knowledge on AD~\cite{Jack_dynamic_biomarkers}.

\section{Conclusion}

In this paper, we addressed three major limitations of CNN-based methods by introducing a novel interpretable, generalizable and accurate deep grading framework. First, deep grading offers a meaningful visualization of the disease progression. Second, we proposed a collective artificial intelligence strategy to improve the generalization of our DG strategy. Experimental results showed a gain for both SVM and GCN in all tasks using this strategy. Finally, we proposed to use a graph-based modeling to better capture AD signature using both inter-subject similarity and intra-subject variability. Based on that, our DG method showed state-of-the-art performance in both AD diagnosis and prognosis.

{\small
\subsubsection*{Acknowledgments}
This work benefited from the support of the project DeepvolBrain of the French National Research Agency (ANR-18-CE45-0013). This study was achieved within the context of the Laboratory of Excellence TRAIL ANR-10-LABX-57 for the BigDataBrain project. Moreover, we thank the Investments for the future Program IdEx Bordeaux (ANR-10-IDEX-03-02), the French Ministry of Education and Research, and the CNRS for DeepMultiBrain project.
}

\newpage
\appendix

\bibliographystyle{splncs04}
\bibliography{references.bib}

\end{document}